\documentclass[aps,showpacs,onecolumn,11pt]{revtex4}

\usepackage{t1enc}
\usepackage[final]{graphicx}
\usepackage{graphicx}
\usepackage{epsfig}

\begin{document}

%\begin{frontmatter}

\title{Distributions of switching times of single-domain particles using a time quantified Monte Carlo method}

\author{Orlando V. Billoni}
\email{billoni@famaf.unc.edu.ar}
\altaffiliation{CONICET, Argentina}
\affiliation{Facultad de Matem\'atica, Astronom\'\i a y F\'\i sica, Universidad Nacional de C\'ordoba, Ciudad Universitaria, 5000 C\'ordoba, Argentina}

\author{Daniel A. Stariolo}
\altaffiliation{Research Associate of the Abdus Salam International Centre for Theoretical Physics,  Trieste, Italy.}
\affiliation{Departamento de F\'{\i}sica,  Universidade Federal do Rio Grande do Sul CP 15051, 91501--979, Porto Alegre, Brazil.}

\date{\today}

\begin{abstract}

Using a time quantified  Monte Carlo scheme we performed  simulations of the switching time distribution of single
mono-domain particles in the Stoner-Wohlfarth approximation. We considered uniaxial anisotropy  and different conditions
for the external applied field. The results obtained show  the switching time distribution can be well described by two
relaxation times, either when the applied field is parallel to the easy axis or for an oblique external field and a
larger damping constant.
We found that in  the low barrier limit these relaxation times are in very good agreement with analytical results
obtained from solutions of the Fokker-Planck equation related to this problem.
When the damping is small and the applied field is oblique the shape of the distribution curves 
shows several peaks and resonance effects.

\end{abstract}


% 75.40.Mg      Numerical simulation studies
% 75.50.Tt      Fine-particle systems; nanocrystalline materials 
%\KEY  Fine-particle systems \sep Monte Carlo simulations \sep Switching time distributions.
%\PACS 75.40.Mg \sep 75.50.Tt 

%\end{frontmatter}

\maketitle

\section{Introduction}
The study of magnetic nanoparticles is very interesting  from the theoretical and experimental point of
view, in addition to its  important technological applications in magnetic data storage.
Magnetic media for data storage  are composed of tiny magnetic grains, like magnetic particles, which have
to be flipped in the magnetic recording process. Increasing the storage density implies a reduction
in the size  of the magnetic grains. If the grains are too small they  lose thermal stability reaching the
so called {\it superparamagnetic limit}. The stability problem  can be overcome by increasing the anisotropy
of the grains but then higher fields (which are  difficult  to reach in current devices) are needed to switch
the magnetization in the recording process.
In this sense a  challenging issue about magnetization reversal arises  which is how to have short reversal times,
keeping the switching fields in small values \cite{Sun05PRB,Sun06PRL} and controlling the effects of thermal
fluctuations.
The thermal stability problem of ferromagnetic particles was first studied by  N\'eel and then by
Brown \cite{Neel49AG,Brown63PR,Brown79IEEE}  in the single spin approximation.  Experimental studies for the
switching time in a single magnetic nanoparticle  \cite{Wernsdorfer97PRL,Wernsdorfer97bPRL,Thirion03N}
suggest in several cases the correctness of the N\'eel-Brown approximation.

In his  work Brown \cite{Brown63PR}  derived the Fokker-Planck equation (FPE) for an assembly of particles using the
stochastic Landau-Lifshitz-Gilbert dynamics (sLLG)  and then obtained analytical expressions for the lowest eigenvalue
of the Sturm-Liouville problem associated with the FPE  in  asymptotic cases. The lowest eigenvalue of the FPE is
associated to the longer relaxation time which in several regimes is supposed to control the  switching  process.
Since then several works have been done to obtain  the lowest eigenvalue of FPE, with applied fields parallel
\cite{Aharoni69PR}  and oblique  \cite{Coffey95PRB,Coffey98PRB} to the easy axis. While in the past the
attention was focused on the long time regime of the switching process, with the growing complexity of experimental
devices and the need of ever smaller switching times, it turns to be desirable to have a knowledge of the whole
process, from the earliest times to the longest ones, in order to explore alternative mechanisms for
magnetization switching.

Since  the FPE cannot  be solved analytically except in limiting cases,   the short
relaxation times in the early stages of the switching process has to be studied by numerical integration 
of the sLLG \cite{Grinstein05PRB} equation, or solving numerically the FPE.
For the intermediate time scales numerical integration of Langevin dynamics is not useful since the small time step  
needed in  the numerical integration does not allow to extend the simulation  beyond the range of nanoseconds.
In order to overcome this problem  time quantified Monte Carlo  methods (TQMC)  were  
developed \cite{Nowak00PRL,Smirnov-Rueda00JAP,Chubykalo02JMMM,Chubykalo03PRB} that allow  to extend
the time span with a considerable  gain in computational and programming efforts.   
In addition, Chubykalo {\it et al.} \cite{Chubykalo03PRB} have also proved that these methods may be useful 
in the low damping condition,  whenever the applied field is parallel to the anisotropy easy axis. However the method 
fails  under an oblique applied field  and low damping condition where precessional movement cannot be neglected.
Recently, Cheng {\it et al.} \cite{Cheng06PRL}  mapped  the sLLG dynamics and the Monte Carlo scheme
through a Fokker-Planck approach introducing the precessional step in the TQMC simulations. With this
improvement in principle  TQMC simulations could be used also in the short times scales irrespectively of the 
configuration of the applied field and the value of the damping constant. However this improvement in the accuraccy
is in detriment of the extent of time of the simulation. The situation is even worse when the system consists of
 a set of interaccting particles \cite{Cheng06PRB}. 
Nevertheless, even in these cases the TQMC scheme seems a better alternative
than solving numerically the sLLG equation. The main advantage is a far simpler implementation, whithout the need to
control the stability of the algorithm as when solving a differential equation numerically. The second advantage 
is that, even though the gain in the time span simulated relative to the sLLG equation is certainly not as large as in the
strong damping regime where precession can be ignored, the TQMC scheme allows to adjust the time step in a less
constrained manner than the sLLG equation, resulting in a real gain in the total times that can be reached by the
simulation. The extent of this gain depends on the particular problem considered.

In this work we used this approach to obtain the distribution of switching times in order to explore the incidence 
of short relaxation times in the switching process as a function of the damping and for different applied fields. 
We found that for external fields parallel to the easy axis the Brown approach works very well in practically the whole 
time span. The numerical results are extremely well described with only the two largest relaxation times from the solution 
of the Fokker-Planck equation. When the external field is oblique there are no analytical solutions available, except for
the largest time scale. In this case, we found that two relaxation times are enough to describe the 
distribution of times when
the damping is high. When the damping is low, the switching mechanism is dominated by precession, and the short time
behavior is more complex, showing several peaks, which reflect the presence of different resonance frequencies. 

\section{Theoretical  Background}
A single-domain particle can be modeled in the Stoner-Wohlfarth (SW) approximation  where the magnetic
state of the particle is described  by a single magnetic moment $\vec{m}$ whose strength is  equal to
the total magnetic moment of the particle $|\vec{m}|=M_s v$. Here $M_s$ is the magnetization of saturation of
the particle and $v$ is the particle volume.
In the SW model the  energy density $V$ of a particle with uniaxial anisotropy under an external applied field
is expressed as:

\begin{equation}\label{eq1}
-\beta V= \alpha [(\vec{n} \cdot \vec{s})^2 + 2\vec{h} \cdot \vec{s}],
\end{equation}

\noindent here $\beta=v/k_B T$,  $\vec{s}=\vec{M}/M_s$ and $\vec{n}$ are unit vectors defining the  magnetization
and the easy axis direction, respectively. The applied field $\vec{h}$ is expressed in reduced units $h=H/H_k$,
with $H_k=2K/\mu_0 M_s$  being the
anisotropy field,  $\alpha=K v/k_B T$ is a dimensionless constant where $K$ is the anisotropy constant.
In a classical approximation, the dynamics of a reduced magnetic moment $\vec s$ under thermal fluctuations
is modeled by the stochastic Landau-Lifshitz-Gilbert (sLLG) equation,

\begin{equation}\label{eq2}
\frac{d \vec{s}}{dt}= \frac{\gamma (\mu_0 H_k)}{1+a^2}\, \vec{s} \times [ (\vec{h}_{eff}+\vec{h}_{fl}) -  \\
a\, \vec{s} \times (\vec{h}_{eff}+\vec{h}_{fl})],
\end{equation}

\noindent where $\gamma$ is the gyromagnetic ratio,  $a=\eta \gamma M_s $ is a dimensionless damping coefficient
and $\eta$ is the damping coefficient in Gilbert's equation. The effective field  $\vec{h}_{eff}$ is given by the
particle energy gradient,

\begin{equation}\label{eq3}
\vec{h}_{eff}=- \frac{1}{2K}\frac{\partial V }{\partial \vec{s}}.
\end{equation}

\noindent The stochastic fluctuating field $\vec{h}_{fl}$ is assumed as a Gaussian stochastic process with the following
statistical properties:

\begin{equation}\label{eq4}
<{h}_{fl,i}> =0,\,\,\,\, <h_{fl,i}(t)\,h_{fl,j}(s)>=\mu\, \delta_{i,j}\, \delta(t-s),
\end{equation}

\noindent where $i$ and $j$ stand for the cartesian components.

In spherical coordinates, the FPE for the probability $W(\theta,\phi,t)$ of finding
one the magnetic moment at time $t$ within a  solid  angle $d \Omega$  is given by
\cite{Brown63PR,Brown79IEEE}:

\begin{eqnarray}\label{eq5}
\frac{\partial W}{\partial t} = \frac{1}{2 \tau_N}\nabla^2 W + ab \nabla ^2 V \, W
+\frac{b}{\sin(\theta)} \left( \frac{\partial V}{\partial \theta}\frac{\partial W}{\partial \phi}
-\frac{\partial V}{\partial \phi}\frac{\partial W}{\partial \theta}\right) \\ \nonumber
+ab\left(\frac{\partial V}{\partial \theta}\frac{\partial W}{\partial \theta} +\frac{1}{\sin(\theta)^2}
+\frac{\partial V}{\partial \phi}\frac{\partial W}{\partial \phi}  \right).
\end{eqnarray}

\noindent  where the N\'eel time  $\tau_N^{-1}= \mu \gamma^2(1+a^2)$  is a characteristic
diffusional time and $b=\gamma/(1+a^2) M_s$.
On the other hand,  in order to satisfy the equilibrium statistical properties in the stationary regime,

\begin{equation}\label{eq11}
\tau_N =  \alpha \frac{(1+a^2)}{a} \frac{1}{ (\gamma \mu_0 H_K)}.
\end{equation}

\noindent  In general solutions of (\ref{eq5}) can not be found analytically. However, the relaxation of
any initial probability state can be formally described by a sum of exponentials:

\begin{equation}\label{eq11b}
W(\theta,\phi) =  W_0 + \sum_{n=1}^{\infty}A_n\,F_n(\theta,\phi)\, \exp(-t/\tau_i),
\end{equation}

\noindent where $\tau_i$ are related to the eigenvalues of the Sturm-Liouville associated
problem \cite{Brown63PR,Aharoni69PR,Coffey98PRB} according to

\begin{equation}\label{eq12}
\tau_i = \frac{2 \tau_N}{\lambda_i}.
\end{equation}

\noindent Besides the diffusional N\'eel time, the other characteristic time scales in this problem
are the precessional time  $\tau_p=(\gamma \mu_0 H_K/(1+a^2))^{-1}$ and the damping time $\tau_K=  \tau_p/a$
\cite{Garcia-Palacios98PRB}. Then,  eq. (\ref{eq12}) expressed in terms of the damping times becomes
$\tau_i[\tau_K] = \frac{2 \alpha}{\lambda_i}$.

\section{Monte Carlo simulations}

In our simulations we use a hybrid Monte Carlo  method \cite{Nowak00PRL,Cheng06PRL} that emulates the
stochastic dynamics of the LLG equation. This method combines  the Metropolis or heat bath MC scheme with a random
displacement of the
magnetic moment within a cone \cite{Nowak00PRL} and a precessional spin motion. The random  displacement is obtained
by adding to the normalized magnetic moment  a random vector uniformly distributed within an sphere of radius $R \ll 1$
and then normalizing the resulting vector again, while the magnitude and direction of the precessional motion is given by

\begin{equation}\label{eq14}
\Delta \vec{s}   =  -\Phi \  \vec{s} \times \vec{h}_{eff},
\end{equation}

\noindent where $\Phi=\frac{\alpha}{10 a}R^2$. In this MC scheme the magnetic moment update is  chosen with equal
probability between a precessional  step and a random displacement. The acceptance rate $A(\beta \Delta V)$ of the
random motion is based in the heat bath procedure,

\begin{equation}\label{eq15}
A(\beta \Delta V)    = 1/[1+\exp(\beta \Delta V)].  
\end{equation} %

\noindent By means of a detailed comparison between the Fokker-Planck equation representing the MC stochastic dynamics
and the corresponding equation associated with the LLG micromagnetic equation, it is possible to obtain a very accurate
mapping between Monte Carlo Steps and the real time scale from the LLG equation~\cite{Cheng06PRL}
through the following relation:

% Real time MCS scaling   I
\begin{equation}\label{eq16}
\Delta t [\tau_K]  =   \frac{\alpha}{20} R^2 \Delta t [MCS] ,
\end{equation}
where the real time scale is expressed in units of the damping characteristic time $\tau_K$.

\section{Results}

We have performed numerical evaluations of the  switching time distribution with the following protocol: we start with
the magnetic moment
pointing in the easy axis  direction, in our case the $z$ axis, and then we apply an inverse magnetic field of different strengths
and direction. The switching time  is defined as the time required  for the z component of the magnetic moment to change its sign.
Since we want to  compare  our results with analytic predictions from the Fokker-Planck equation~\ref{eq5}, we count every time the magnetic moment crosses the
equatorial line, i.e., we compute the whole probability that the magnetic moment attains an angle $\theta=\pi /2$ in a
given time $t$. Otherwise we would be computing the first passage time, for which there are still less analytic results
available. We performed $10^6$ realizations to obtain  the switching time
distribution $P(t)$. Since $P(t) \propto \int_0^{2\pi} W(\theta=\pi/2,\phi,t)\, d\phi$,  it has the same relaxational behavior than $W(\theta,t)$.

In order to test the confidence of our results we simulated switching times distribution for the low barrier  limit
($\alpha < 1$), with  the applied field parallel to the easy axis. In this case the two smallest eigenvalues
are given by \cite{Brown63PR},

% Eigenvalues in low barrier limit
\begin{eqnarray}\label{eq17}
\lambda_1 = 2 - \frac{4}{5}(1-h^2\alpha)\alpha + \frac{96}{875}\alpha^2 + \vartheta(\alpha^3) \\
\lambda_2 = 6 - \frac{4}{7}(1-h^2\alpha)\alpha + \frac{64}{343}\alpha^2 + \vartheta(\alpha^3). \nonumber
\end{eqnarray}

\noindent From  equations (\ref{eq12}) and (\ref{eq16}), and considering that a time of $3\tau_1$ is enough to
obtain a complete distribution curve, the number of MCS that should be used for each realization is:

\begin{equation}\label{eq18}
\Delta t [MCS] =    \frac{60}{R^2 \lambda_i }.
\end{equation}

\begin{figure}%[ht]
\includegraphics*[width=9cm,angle=-90]{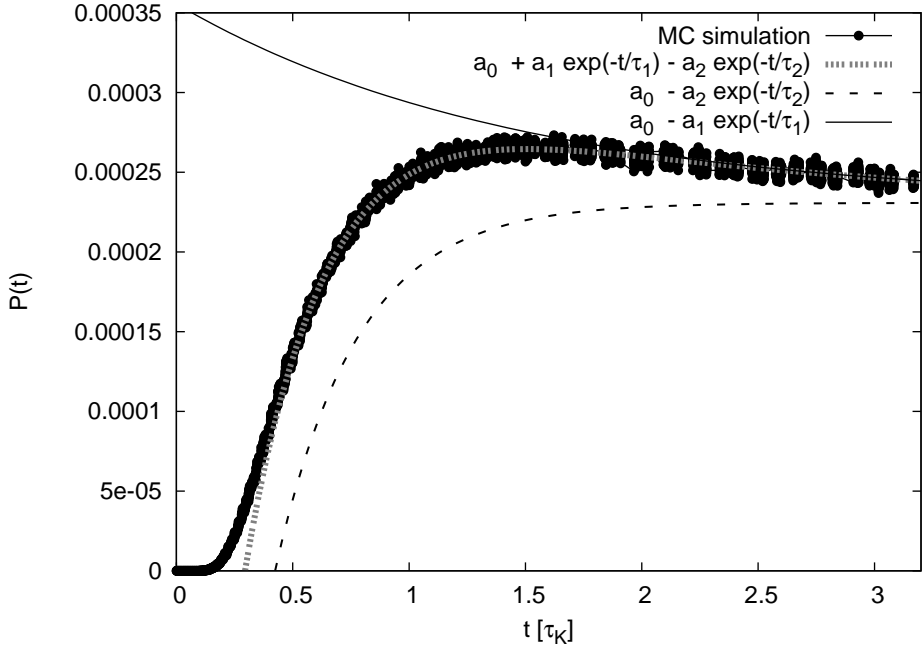}
\caption{\label{fig0} Switching time distribution in the low barrier case $ \alpha=1 $ and parallel applied field  $h=0.292$,
 here the damping constant is $a=1$. The relaxation times used in the fitting  $\tau_1=7.71$ and $\tau_2=1.79$ are obtained from eqs. 
(\ref{eq12}) and (\ref{eq17}).}
\end{figure}

\noindent In the simulations we used $R=0.03$, which is a good compromise between accuracy and efficiency.
In figure \ref{fig0} we present the switching time distribution for
an external field $h=0.292$ parallel to the easy axis and low energy barrier $\alpha=1$.
At short times the probability that the particle switches is zero
since there is a minimal time necessary to surmount the barrier. Except for the very short times, the distribution is
well   fitted using two exponentials
$P(t)=a_0+a_1 \exp(-t/\tau_1)+a_2\exp(-t/\tau_2)$  where the relaxation times  $\tau_1$ and $\tau_2$ are obtained
through eq. (\ref{eq12}) using the eigenvalues given in eqs. (\ref{eq17}).
We can see that two relaxation times are enough to obtain a good fit of the switching time distribution in a wide range and
a very good agreement for the relaxation times obtained through the FPE.
The  switching time distribution at large times is finite because  the energy barrier is small and the particle attains
thermodynamic equilibrium with a finite probability of being at $\theta=\pi /2$.

%FIGURA 1

\begin{figure}%[ht]
\includegraphics*[width=9cm,angle=-90]{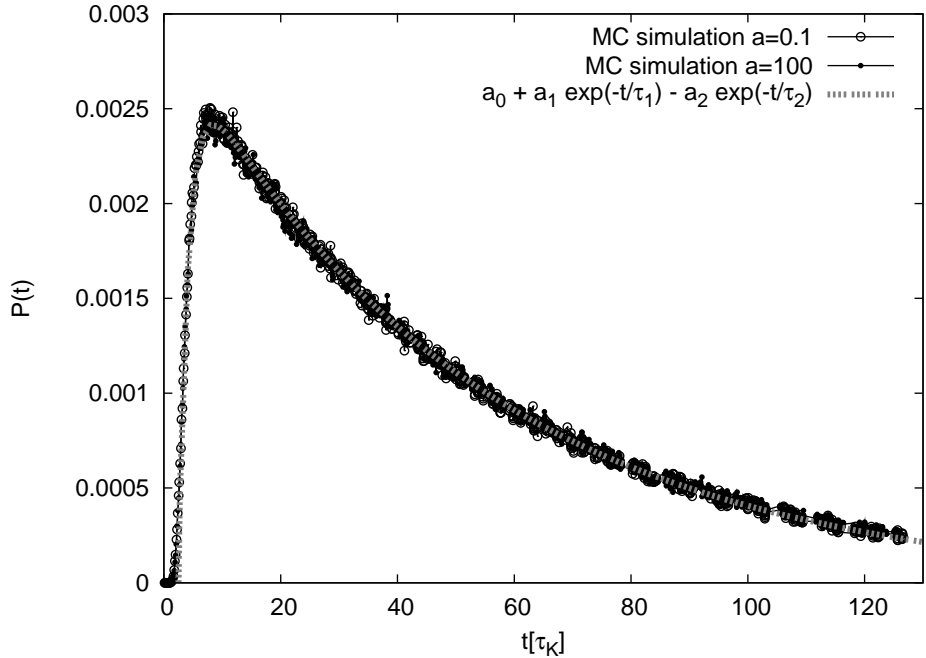}
\caption{\label{fig1}  High barrier switching time distribution with $\alpha=10$ for two damping constants $a=0.1$ and $100$ and parallel 
applied field $h=0.4$. The longer relaxation time  $\tau_1=52.0$ is obtained from table I \cite{Coffey98PRB}. The 
secondary relaxation 
time is obtained from fitting:  $\tau_2=1.72$ and the mean time $t_m=42.5$.}
\end{figure}

In fig.  \ref{fig1}, the energy barrier is increased ($\alpha=10$) keeping the field parallel, with $h=0.4$. In this case
the switching time distribution
has a peak before  relaxing  to zero. In this case the distribution goes to zero at large times since the magnetic moment
gets trapped in the
deepest minimum A similar behavior is found  by integrating the sLLG equation \cite{Grinstein05PRB}.
Like in fig. \ref{fig0},  this curve is well fitted by two relaxation times. The longer time $\tau_1$ used in this fitting
corresponds to  results of Coffey {\it et al.} \cite{Coffey98PRB}. The secondary relaxation time $\tau_2$ ,
which is supposed to  be related to the second eigenvalue is much lower than $\tau_1$ and is important in the very first
stages
of the relaxation, having little influence in the switching mean time $<t>$.
We also show in  fig. \ref{fig1} two curves corresponding to small and large damping constants $a=0.1,\, 100$.
These curves are
indistinguishable  within the statistical error, when plotted in units of the damping time $\tau_K$. This is because if
the applied  field  is parallel to the easy axis, then the potential energy has azimuthal symmetry and the  precessional
motion, which is important for small damping,  has no influence in crossing the barrier. We will see below that
a different situation is observed if the applied field is oblique. However, from the point of view  of the Monte Carlo
simulation, the damping constant has
a critical influence since the precessional factor $\Phi$ has to be kept at small values in order to
correctly follow the trajectory. Then, reducing the damping  constant implies a
reduction of $R$ in eq. (\ref{eq14}), and the number of MCS required in the simulation notably increase
(see eq. (\ref{eq18})).

%FIGURA 2

\begin{figure}%[ht]
\includegraphics*[width=9cm,angle=-90]{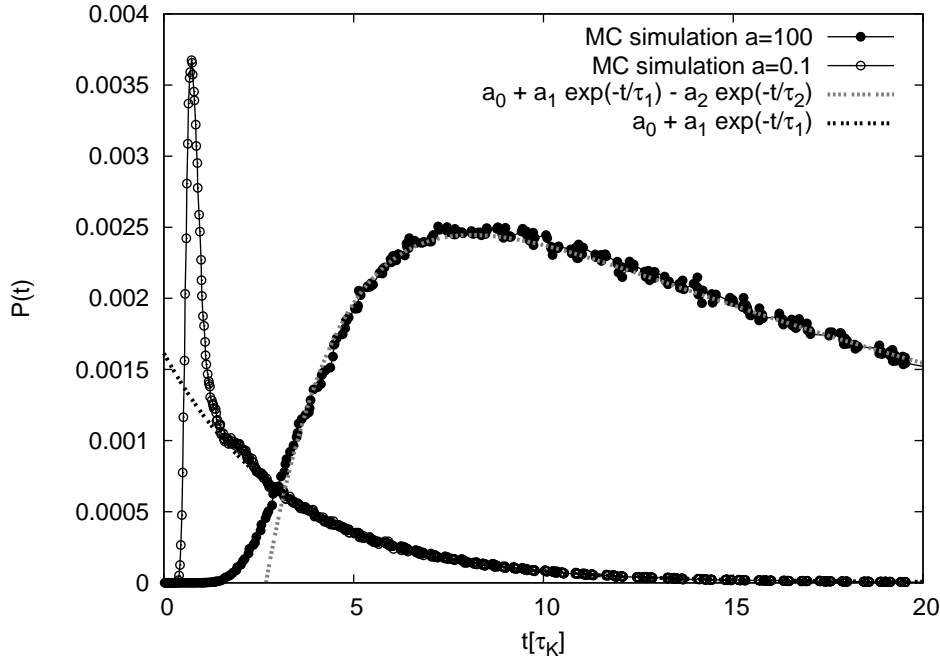}
\caption{\label{fig2} High energy barrier $\alpha = 10$ under an oblique applied field $\phi=\pi/4, \, h=0.4$ for two different values of 
the damping constant $a=0.1, \,100$. It can be seen that in the larger damping case  the distribution is well described by two relaxation
times whereas in the low damping case a secondary peak appears.}
\end{figure}

Figure \ref{fig2} is similar to fig. \ref{fig1}, $h=0.4$, but now the  applied field is at an angle of $\pi/4$ with
respect to
easy axis. The damping constant has now influence on the switching behavior, this is due to the coupling between the
longitudinal
and normal modes \cite{Kalmykov04JAP}. When the applied field is oblique the particle crosses the barrier through
a saddle point
and the precessional motion influences the way the energy landscape is explored. If the damping constant is high the particle
tends to relax to the nearest minimum and only through thermal fluctuations the saddle point can be found, wheres if the damping
constant is kept small, precessional motion affects the switching process. Note that a shoulder is present after
the first peak in the case of small damping.

% FIGURA 3

\begin{figure}%[ht]
\includegraphics*[width=9cm,angle=-90]{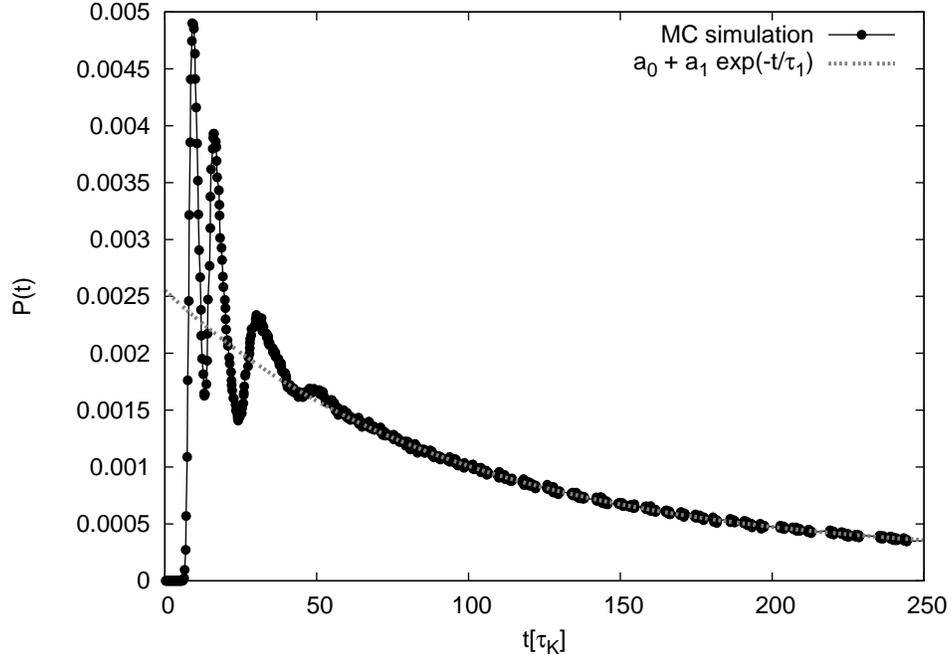}
\caption{\label{fig3} Switching time distribution for a very low damping value $a=0.01$. Like in fig. \ref{fig2}  here 
$\alpha=10$, $\phi=\pi/4$ and $h=0.4$. We can see at least four peaks and an exponential relaxing behavior at 
the longest times.}
\end{figure}

\begin{figure}%[ht]
\includegraphics*[width=9cm,angle=-90]{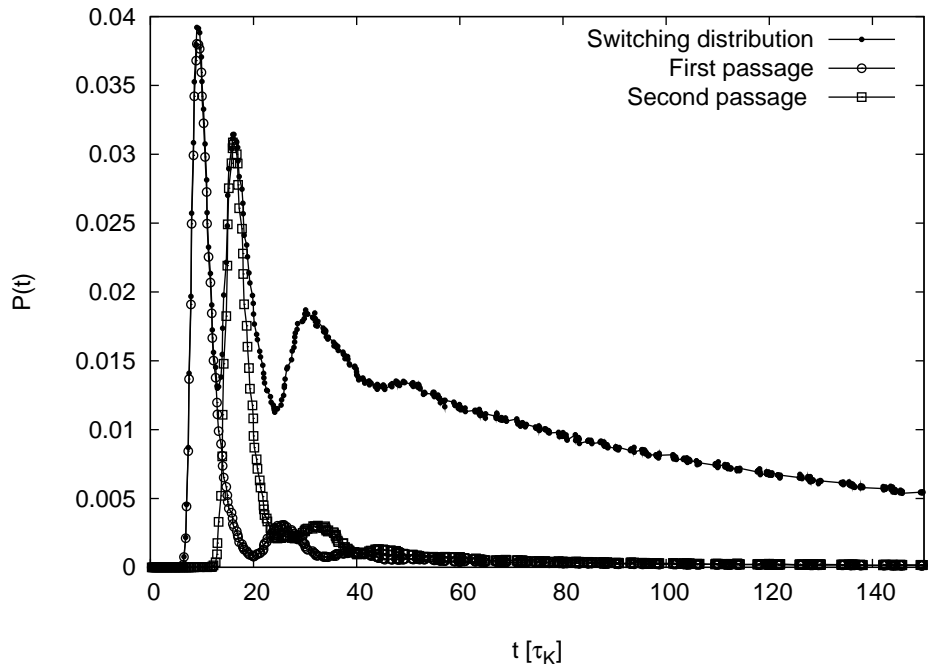}
\caption{\label{fig4} Switching time distribution of fig. \ref{fig3} together with the first and second passage time distribution.
Clearly the first and second peak of the total distribution correspond to the first a second passage, respectively.}
\end{figure}

In fig. \ref{fig3} the damping  constant is decreased even more to $a=0.01$. Now new peaks appear in the switching
time distribution, which shows resonance-like effects. If the starting energy is similar to the energy of the saddle point
and the damping constant is low enough, the particle does not relax quickly and keeps its energy nearly constant
for a long time, letting the magnetic moment to cross the equatorial line more than one time. This behavior is
analyzed in more detail in figure \ref{fig4}, where the curve of
fig. \ref{fig3} is plotted together with the distribution of the first and second passage times. The figure shows that
the first and second peaks in the switching time distribution correspond to the main peaks of the first and second
passage times distributions, respectively. These distributions show also secondary peaks, which probably correspond to
different paths of switching in the energy landscape. From the figure is also clear that the first passage time
probability goes to zero at $t/\tau_K \approx 70$, while the whole distribution stays finite until much longer times.
This fact means that the magnetic moment can go back and forth across the saddle point and the magnetic moment keeps
switching between the basins of the two minima during a long time. Although the barrier is high, the small damping makes
the magnetic moment to follow a long trajectory before settling in the final state.

% FIGURA 4

% FIGURE 5

\section{Summary}
In summary, by performing Monte Carlo simulations including a precessional step, we have obtained switching time 
distributions of magnetic particles in the Stoner-Wolfarth limit, 
for different configurations of the applied field, different values
of the damping constant and different heights of the energy barriers. 
We conclude that if the damping constant is high enough the distributions are well described by two relaxation
times associated with the eigenvalues of the Fokker-Planck solutions of the corresponding Landau-Lifshitz-Gilbert 
dynamics, irrespectively of the configuration of the applied field. 
If the damping constant takes small values and the applied field is oblique to the anisotropy axis, the distributions 
show resonance effects, evidencing the importance of the precessional motion in the inversion mechanism. 
The present Monte Carlo algorithm allows to study these precessional effects in detail, without the need to solve the 
LLG equations. We showed that the first two peaks in the distribution functions correspond to the first and second 
passage times of the magnetization through the equator. 
In all cases the characteristic inversion time is given by the smallest eigenvalue of the FPE, 
while in the cases with strong damping the whole distribution can be very well described by only two relaxation times.

\section{Acknowledgments}
This work was partially supported by grants from FAPERGS and CNPq (Brazil),
and by  grants from CONICET  and SECYT/UNC (Argentina).

%\bibliographystyle{elsart-num}
%\bibliography{switching}

%\appendix 
%\section{}

\end{document}